\documentclass[manuscript]{aastex}

\newcommand{\ltsimeq}{\raisebox{-0.6ex}{$\,\stackrel
        {\raisebox{-.2ex}{$\textstyle <$}}{\sim}\,$}}
\newcommand{\gtsimeq}{\raisebox{-0.6ex}{$\,\stackrel
        {\raisebox{-.2ex}{$\textstyle >$}}{\sim}\,$}}

\slugcomment{AJ ver \today}

\received{}
\revised{}
\accepted{}

\shorttitle{Albedos of Small Hilda Group Asteroids}
\shortauthors{Ryan \& Woodward}

\begin{document}


\title{Albedos of Small Hilda Group Asteroids as Revealed by Spitzer}


\author{
ERIN LEE RYAN, CHARLES E. WOODWARD
}
\affil{Department of Astronomy, School of Physics and Astronomy, 116 
Church Street, S.~E., University of Minnesota, Minneapolis, MN 55455,\\ 
\it{ryan@astro.umn.edu, chelsea@astro.umn.edu}}

\begin{abstract}
We present thermal 24 $\mu$m observations from the \textit{Spitzer Space Telescope} of 62 Hilda asteroid group members with diameters ranging from 3 to 12 kilometers. Measurements of the thermal emission when combined with reported absolute magnitudes allow us to constrain the albedo and diameter of each object. From our \textit{Spitzer} sample, we find the mean geometric albedo, $p_{V} =$ 0.07 $\pm$ 0.05 for small (D $<$ 10 km) Hilda group asteroids. This value of $p_{V}$ is greater than and spans a larger range in albedo space than the mean albedo of large (D $\gtrsim$ 10 km) Hilda group asteroids which is  $p_{V} =$ 0.04 $\pm$ 0.01. Though this difference may be attributed to space weathering, the small Hilda group population reportedly displays greater taxonomic range from C-, D- and X-type whose albedo distributions are commensurate with the range of determined albedos. We discuss the derived Hilda size-frequency distribution, color-color space, and geometric albedo for our survey sample in the context of the expected migration induced "seeding" of the Hilda asteroid group with outer solar system proto-planetesimals as outlined in the "Nice" formalism.
\end{abstract}



\section{INTRODUCTION}\label{sec:intro}

%% cew reworks this a bit 20101219

Residing in the outer main belt at a distance of $\simeq 4.04$~AU in the 
first-order Jupiter J3:2 mean resonance is an asteroid population with an 
unknown origin, the Hilda asteroid group \citep{gradie89}. Early dynamical 
models suggested that the Hilda group originated field asteroids which were 
captured in gravitational resonances as Jupiter migrated in a sunward 
direction \citep[e.g.,][]{franklin04}. In contrast, more expansive models 
designed to explain the origins of a wide variety of outer solar system 
small bodies families, the `Nice Model' \citep{levison09} argues that 
delivery of proto-Kuiper Belt planetesimals into stable inner solar system 
orbits, such as exhibited by the Hilda asteroid group, occurred populating 
the dynamical families observed at the present epoch. Though the Nice Model 
favors migration from the Kuiper Belt, observational evidence of such small 
body migration is tentative at best. Could the Hilda group asteroids be the 
remnant population arising from the purported migration effects detailed in 
the Nice Model or are the antecedents field asteroids? Answers to this 
question have a direct bearing on the efficiency of early solar system 
dynamical processes outlined in the Nice model. 

A detailed analysis of the colors and albedos of the Hilda population 
yields clues to their origin and can substantiate outcomes described by the 
Nice formalism. Large (diameters $\gtsimeq 10$~km) Hilda asteriods have a 
range of $V-R$ color magnitude, 0.38 to 0.49 \citep{dahlgren98}, which is 
similar to those colors of comets \citep{hainaut_delsanti02}. In particular 
ecliptic comets (ECs) span the same narrow range of $V-R$ \citep{lamy2009} 
as the Hildas, and are thought to have originated as outer solar system 
Centaurs which were subsequently scattered into the inner solar system due 
to gravitational interactions with Jupiter. C-type asteroids in the outer 
asteroid belt also inhabit this range of color space. However, the $V-R$ 
colors of smaller Hildas show greater spectral diversity 
\citep{sloanhildas08} commensurate with the colors of both C- and 
X-asteroid taxonomic type as well as some Kuiper Belt objects 
\citep{gulbis06}. 

The geometric albedos of large Hilda group asteroids from \citet{ryan2010} 
are commensurate with the albedo range for ECs \citep{lamy04} and other icy 
bodies such as main belt comets \citep{hsieh09}. The albedo segregation of 
cold classical Kuiper Belt objects (KBOs) and hot classical KBOs is argued 
to indicate differing origins of these two populations \citep{brucker09}. 
If a Nice model-like migration occurred and the small Hilda group asteroids 
did indeed originate in the proto-Kuiper Belt, we expect that the albedos 
of these objects would be commensurate with the albedos of the low 
inclination, cold classical KBOs. 

Here, we present an analysis of the albedo and diameter of Hilda family 
asteroids extracted from the NASA \textit{Spitzer}\, \citep{werner04} 
infrared (IR) archival database, cross referenced with extant optical 
photometry. In \S\ref{sec:obs}, generation of the asteroid survey sample is 
presented, \S\ref{sec:therm_mods} outlines our thermal model analysis of 
the photometry, \S\ref{sec:disc} discusses the outcomes of our analysis, 
including an estimate of the size-frequency distribution inferred from the 
Hilda asteroid group sample and whether the Hilda group are indeed a 
migrant population, while concluding remarks are summarized in 
\S\ref{sec:concl}. 

\section{ARCHIVAL ANALYSIS}\label{sec:obs}

The photometry discussed herein was obtained from the NASA \textit{Spitzer} 
Archive and represents all Hilda asteroids observed in Program 
Identification number (PID) 40819. These data were obtained in Cycle-4 of 
the \textit{Spitzer} cryogenic mission with the Multiband Imaging 
Photometer instrument \citep[MIPS;][]{rieke04}. The MIPS 24~$\mu$m band 
imager is a $128 \times 128$ pixel Si:As impurity band conduction detector 
with an effective wavelength of 23.68~$\mu$m with a native pixel scale of  
2.49 arcsec $\times$ 2.6 arcsec. All 24~$\mu$m images are diffraction 
limited. All targets in PID 40819 were observed in MIPS Photometry mode; 
data was only obtained in the 24 and 70~$\mu$m channels.  No useful 
70~$\mu$m data exists due to the offset of $\sim$ 12 arcminutes between the 
two fields of view. 
 
Observations utilizing the Compact Source Photometry (CSP) template in this 
PID consist of 14 images each with an exposure of 3~sec in length resulting 
in a total observation time per object of 42~sec. Each observation utilized 
10.55~mins of spacecraft time including overhead. Basic data processing 
including removal of dark current, flat fielding, and flux calibration was 
performed via the automated \textit{Spitzer} pipeline ver. 18.2.0 
\citep{Gordon05} to create basic calibrated data (BCDs). 

We used the MOPEX \citep{mm05} program to obtain photometry on all 14 BCDs 
in an image data stack utilizing point-spread function (PSF) photometry 
while subtracting the median background, and then averaged fluxes and 
calculated errors. Photometry reported in Table~\ref{table:flux_table} is 
the average of the 14 BCDs in the stack. Uncertainties reported in 
Table~\ref{table:flux_table} include the photometric fitting errors 
reported from MOPEX  and the absolute calibration uncertainty of the 
24~$\mu$m channel of order 2\% \citep{engelbracht07} added in quadrature. 

\section{THERMAL MODELS}\label{sec:therm_mods}

To determine the diameters and albedos of the Hilda group asteroid in our 
sample, we have utilized both the Standard Thermal Model 
\citep[STM;][]{lebofsky89} and the Near Earth Asteroid Thermal Model 
\citep[NEATM;][]{harris98}. The STM and the NEATM both rely upon a basic 
radiometric method to determine the diameter and albedo of an asteroid 
\citep[for details see][]{ryan2010}.  Both models assume balance between 
incident radiation and emitted radiation, where the emitted radiation has 
two components; a reflected and a thermal component. The reflected 
component has approximately same spectral energy distribution (SED) as the 
incident radiation; i.e., the reflected component is dominant in the 
optical and peaks in V band commensurate with the spectral region in which 
the sun emits the greatest flux. The reflected asteroid flux is 
proportional to the diameter of the body, $D\rm{(km)}$ and the geometric 
albedo, $p_{V}$. To maintain energy balance the thermal flux is 
proportional to the amount of incident flux which is not reflected; 
$F_{thermal} \propto D^{2}\,(1-p_{v})$. 

However, asteroids do not maintain one single body temperature, T(K), 
rather there is a temperature distribution across the surface which is then 
radiometrically observed in the mid-IR. The STM and the NEATM utilize two 
different assumed temperature distributions to model the total IR flux, 
which then yields the geometric albedo. The temperature distributions 
invoked by each model are expressed as: 

\begin{equation}
T_{STM} (\Omega) =  \left [\frac{(1-A)S_\odot}{0.756 r_{h}^{2} \epsilon 
\sigma} \right]^{\frac{1}{4}} (cos \Omega)^{\frac{1}{4}} \label{eqn:stm_e}
\end{equation}

\noindent and 

\begin{equation}
T_{NEATM}(\phi, \theta)= \left [ \frac{(1-A)S_\odot}{\eta r_{h}^{2} \epsilon 
\sigma} \right]^{\frac{1}{4}}  (cos \phi)^{\frac{1}{4}} (cos \theta)^{\frac{1}{4}} 
\label{eqn:neatm_e}
\end{equation} 

\noindent where the temperature, T is in Kelvin, A is the geometric Bond 
albedo, S$_{\odot}$ is the solar constant (W~m$^{-2}$), r$_{h}$ is the 
heliocentric distance (AU), $\epsilon$ is the emissivity of the object 
 which is assumed to be 0.9, an appropriate value for rock \citep{morrison1973}, $\sigma$ is the Stefan-Boltzmann constant, $\Omega$ is 
the angular distance from the sub-solar point on the asteroid, $\eta$ is 
the beaming parameter, $\phi$ is the latitude, and $\theta$ is longitude of 
the coordinate grid superposed on the asteroid. 

In the NEATM temperature distribution, $\eta$, the beaming parameter is 
utilized as a varying parameter to characterize both shape and thermal 
inertia. In an ideal case where an asteroid is a perfect sphere with zero 
inertia, $\eta= 1.0$. Only one thermal (24~\micron) photometric measurement 
is available from the Hilda asteroid group measurements 
(Table~\ref{table:flux_table}); therefore, NEATM was run with a fixed 
beaming parameter of $\eta = 0.91$. This value of $\eta$ was selected by 
averaging the value of $\eta$ for 23 large Hilda group asteroids observed 
by IRAS and/or MSX derived by \citet{ryan2010}.  In addition we adopted a 
phase slope parameter ($G$) of 0.15 for the purposes of modeling the PID 
40819 asteroid 24~\micron{} photometry. To compute the geometric albedo 
and thus the temperature distribution on the illuminated face of the 
asteroid, one must anchor solutions with optical radiometric measurements. 
We utilized absolute magnitudes ($H$) from the Minor Planet Center 
\footnote{www.cfa.harvard.edu/iau/mpc.html}(MPC) for this purpose. 

All albedo and diameter solutions reported in 
Table~\ref{table:solutions_table} are derived from Monte Carlo modeling. A 
500 data point distribution was created for each object observation such 
that the mean flux was equal to the 24~\micron{} flux measured by MOPEX and 
the standard deviation of the distribution was equivalent to the 
uncertainties in the flux measurement. These flux points were then used in 
conjunction the known orbital parameters and the absolute magnitude (H) to 
produce albedo and diameter fitting results.  Due to the wide width of the 
24~\micron{} channel, a color correction is also required to accurately fit 
the albedo and diameter.  Our implementation of STM and NEATM applies the 
color corrections iteratively, such that a color correction is applied with 
each refinement of the albedo \citep{ryan2010}. Instead of using the 
subsolar temperature for the color correction, we calculate the mean of the 
temperature distribution for the application of the color correction, as 
described in \citet{ryan2010}. With the STM, albedos were computed over the 
range of color corrections varying the temperatures from 70 to 300~K, to 
examine the fidelity of our approach as our color correction methodology 
differs from other described in the literature \citep[e.g.,][]{brucker09, 
stansberry07}. The albedo deviations over the latter temperature range of color 
corrections were $\ltsimeq 2$\% indicating that uncertainties in the 
derived albedo from color corrections are low compared to the uncertainties 
in the photometry. The albedos and diameters listed in 
Table~\ref{table:solutions_table}, are the mean of the 500 Monte Carlo 
solutions for each asteroid. The standard deviation of these solutions is 
taken to be the statistical uncertainty ($\pm$) in the results listed in 
Table~\ref{table:solutions_table}.

\section{RESULT \& DISCUSSION}\label{sec:disc}

\begin{figure}
\epsscale{1.0}
\plotone{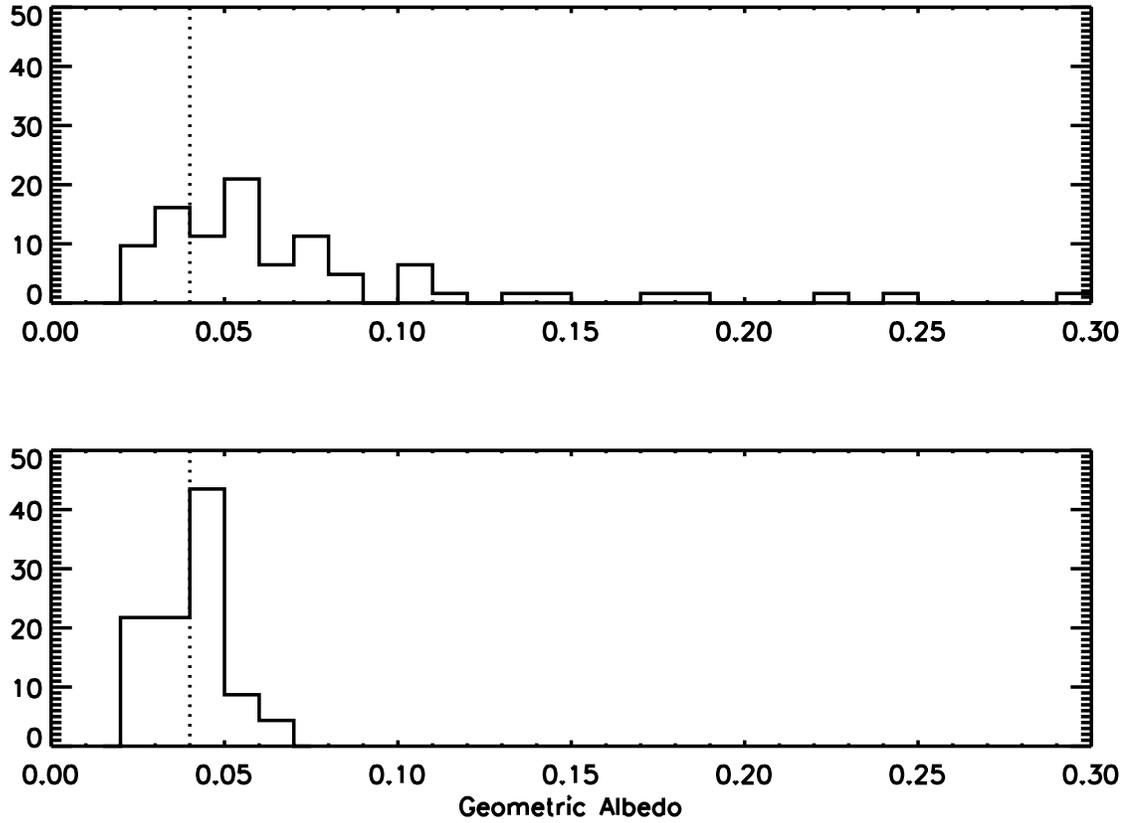}
\caption{Histogram plots of IRAS albedos \citep[derived from][]{ryan2010} from NEATM (bottom panel) and 
\textit{Spitzer} albedos from NEATM (top panel) normalized such that 
the sum of all objects in each sample equals 100.The dashed line indicates the mean albedo of the Hilda 
asteroids detected with IRAS \citep{ryan2010}.
\label{fig:iras_v_spitzer}}
\end{figure}

The distinct differences between the albedo distribution of Hilda asteroid 
group in the IRAS database \citep{ryan2010} as opposed to those observed in 
the MIPS 24~\micron{} survey is clearly illustrated in 
Fig.~\ref{fig:iras_v_spitzer}. The mean NEATM geometric albedo of Hildas 
derived from IRAS photometry is $p_{V}= 0.04 \pm 0.01$ whereas, the mean 
NEATM albedo for asteroids from \textit{Spitzer} MIPS photometry is 
$p_{V}=0.07 \pm 0.05$. These two albedo distributions appear similar; 
however, the distribution for the small asteroids (D~$\ltsimeq 12$~km) 
exhibits a high albedo tail which is not seen in the large (D $> 12$~km) 
asteroid population. 

\begin{figure}
\epsscale{1.0}
\plotone{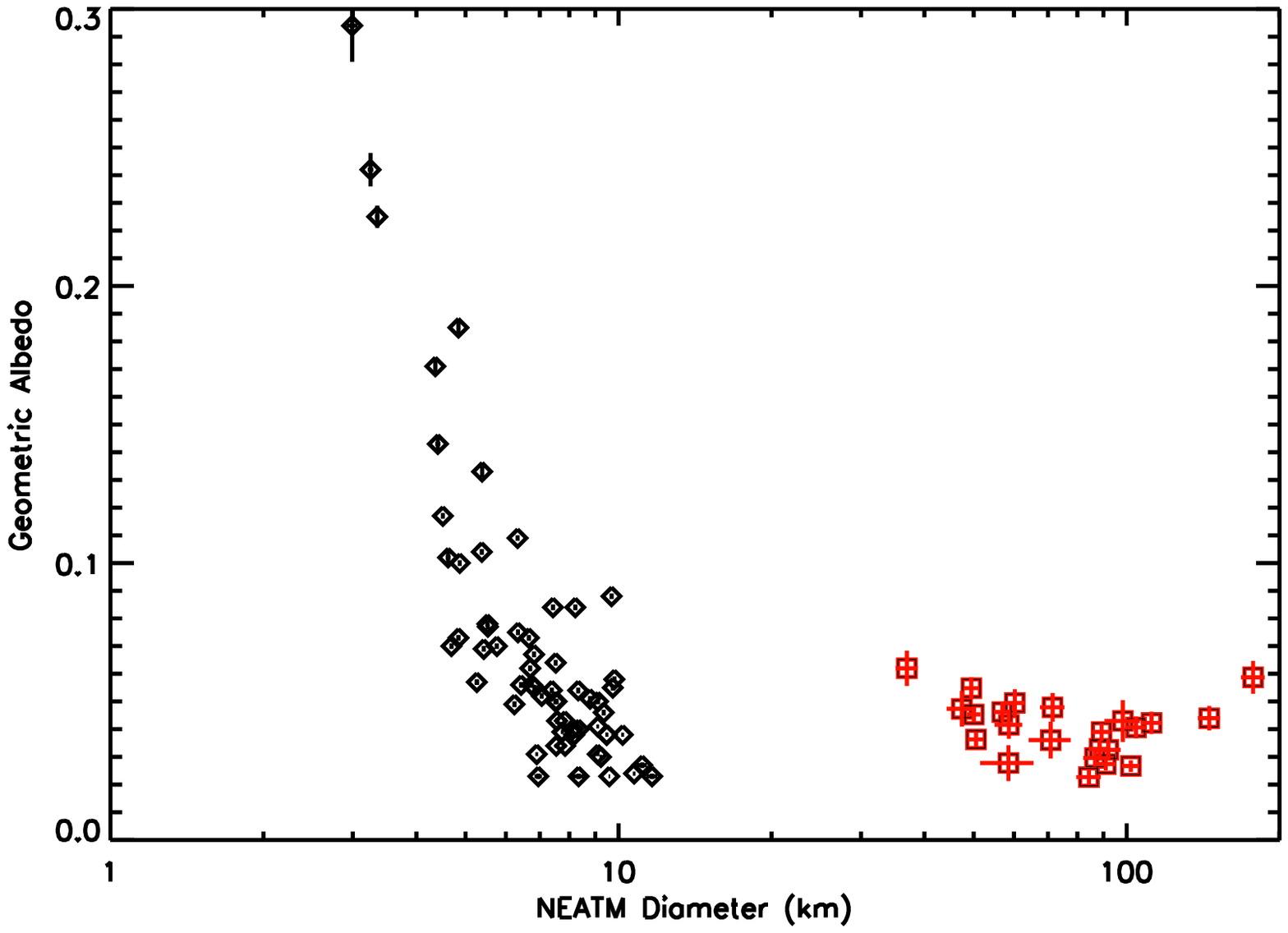}
\caption{Albedo as a function of diameter for Hilda asteroids in 
IRAS (squares) and from \textit{Spitzer} (diamonds) and corresponding 
error bars. 
\label{fig:all_hildas_alb_diam_scatter}}
\end{figure}

\begin{figure}
\epsscale{1.0}
\plotone{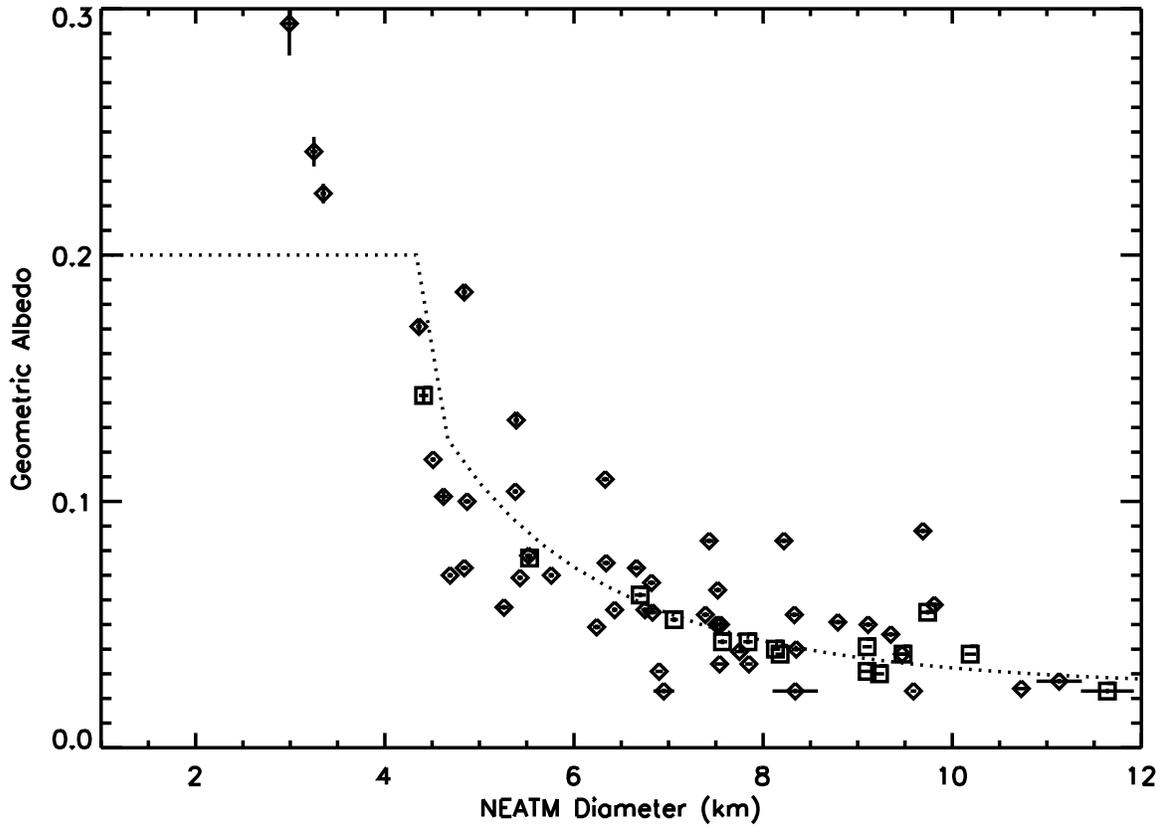}
\caption{Albedo as a function diameter for Hilda asteroids in 
\textit{Spitzer} data. Diamonds denote Hilda `field' asteroids and 
squares are members of the Schubart dynamical family 
(see \S \ref{sec:origins}). The dotted line represents the albedo-diameter 
relation as described in \S \ref{sec:wds_sfd}.
\label{fig:spitzer_alb_diam_scatter}}
\end{figure}

Figures~\ref{fig:all_hildas_alb_diam_scatter} 
and~\ref{fig:spitzer_alb_diam_scatter} illustrate the trend of geometric 
albedo ($p_{V}$) as a function of diameter (in km). Notable is the trend 
that higher albedos are associated with decreasing asteroid diameters. To 
test the significance of this trend, we have calculated the Spearman 
rank-order coefficient \citep[e.g.,][]{meyers2003} for all Hildas with 
known albedos from derived from the IRAS archive \citep{ryan2010} and this 
work, as well as the rank-order coefficient for solely the Hildas from the 
\textit{Spitzer} MIPS survey (Table~\ref{table:solutions_table}). The 
Spearman rank-order coefficient (Spearman $r_{s}$) is commonly defined as 

\begin{equation}
r_{s} = 1 - \left [ \frac{6 \Sigma d_{i}^{2}}{n(n^{2}-1)} \right ],
\label{eqn:spear_r}
\end{equation}

\noindent which assess the statistical dependence between two variables 
assuming that the relationship can be described by a monotonic function. 
The $r_{s}$  for the complete, aggregate Hilda asteroid group data set 
(total population sample $n=85$) is $-0.69$ which corresponds to a 
probability of non-correlation between diameter and albedo of $3.94 \times 
10^{-13}$. The correlation is significant to the 6.3~$\sigma$ level. Using 
only the smaller Hilda population from the \textit{Spitzer} MIPS data 
($n=62$), $r{s} = -0.76$ and the probability of non-correlation is $4.54 
\times 10^{-13}$ with a correlation significance of 5.97~$\sigma$. Whether 
or not the correlation between albedo and diameter may be described by a 
monotonic trend may be influenced by the optical survey completeness. 
However, we assert that the optical detection bias is not the cause of the 
derived albedo increase at small diameters as elucidated in 
\S~\ref{sec:eocmp}. 

\begin{figure}

\plotone{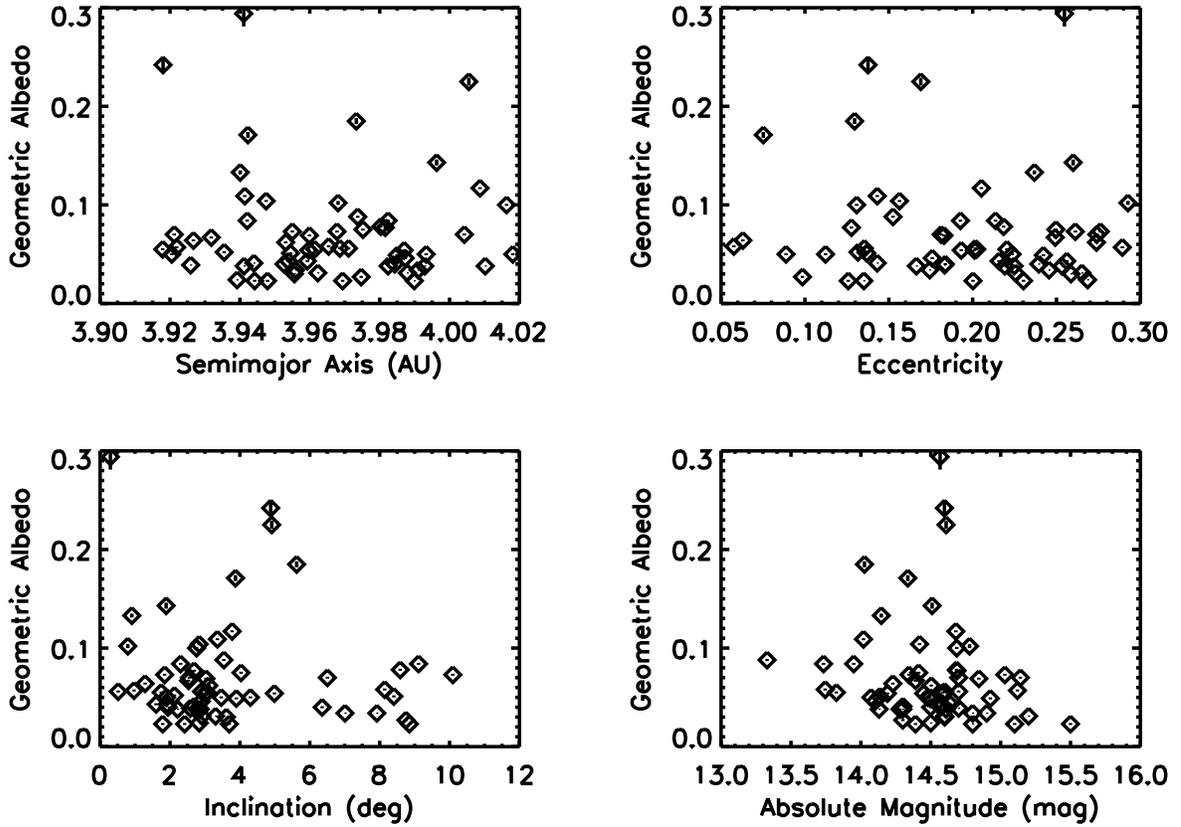}
\caption{\textit{Spitzer} only NEATM derived geometric albedos
$p_{V}$, as function of semi-major axis, eccentricity, orbital 
inclination and absolute magnitude (H) for members of the Hilda
asteroid group deteced in the MIPS 24~\micron{} survey.
\label{fig:spitzer_aeiH}}
\end{figure}

\begin{figure}
\epsscale{0.75}
\plotone{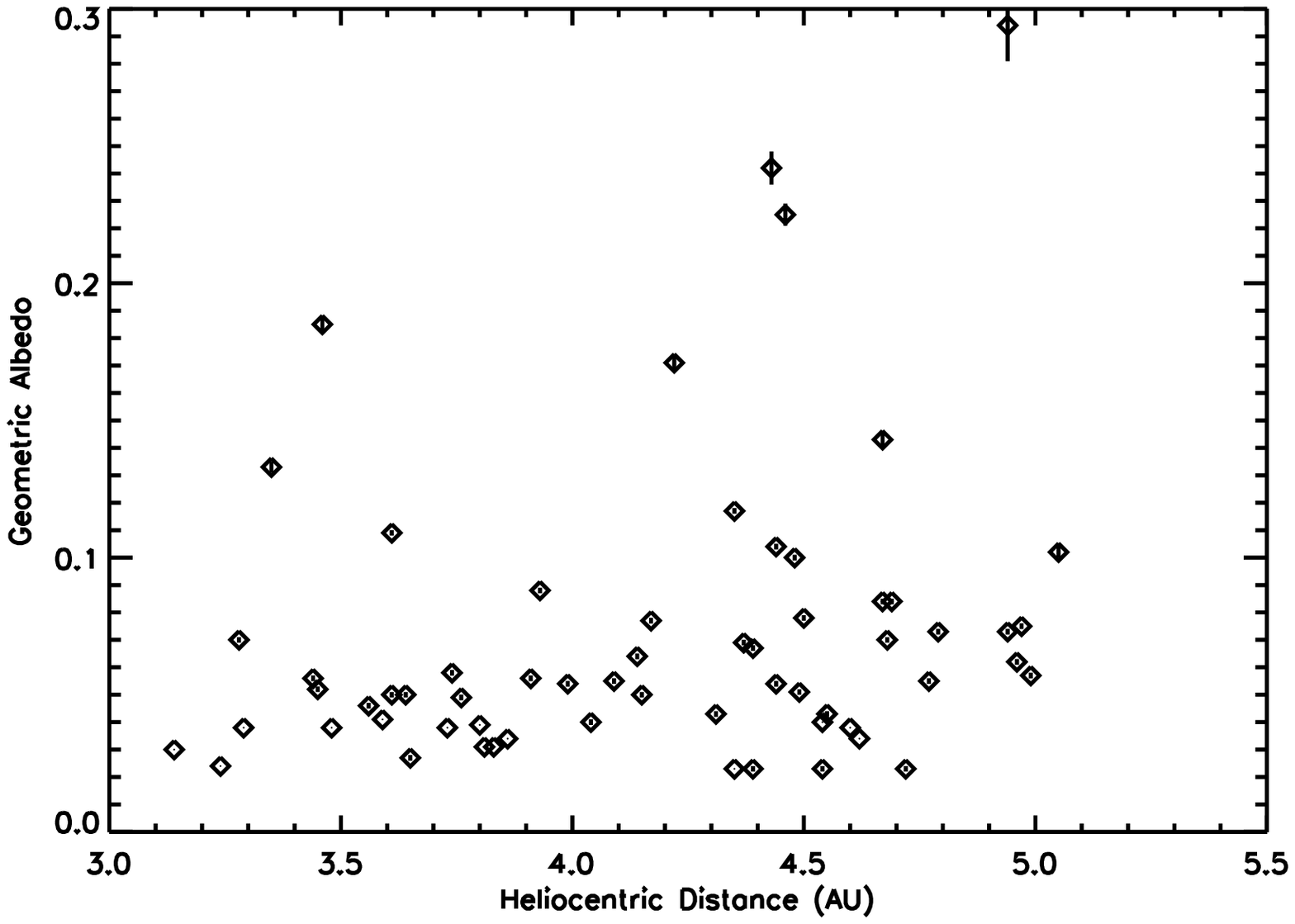}
\plotone{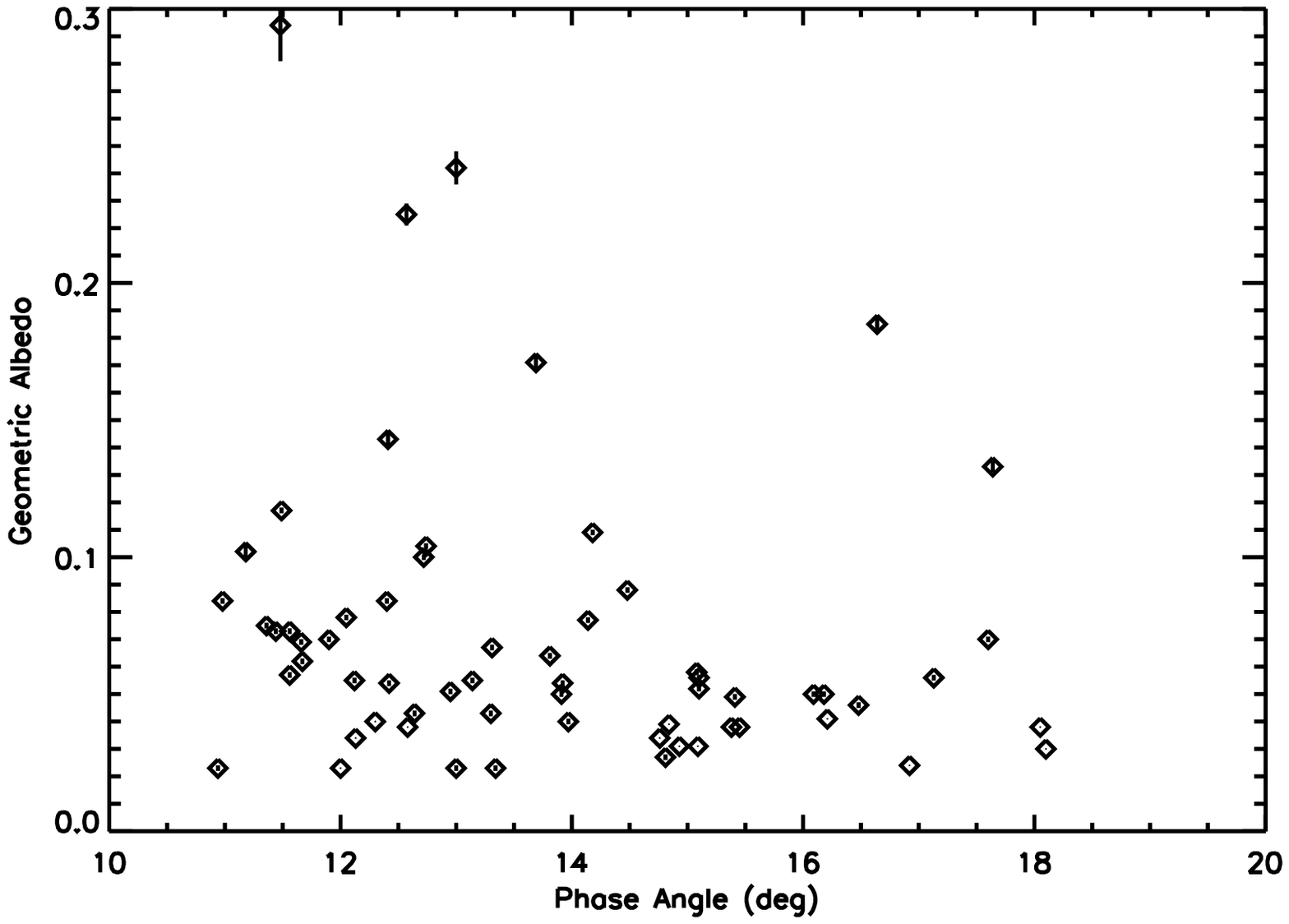}
\caption{\textit{Spitzer} only NEATM derived albedos as a function of 
heliocentric distance and phase angle  for members of the Hilda
asteroid group deteced in the MIPS 24~\micron{} survey.
\label{fig:spitzer_alb_rh_alpha}}
\end{figure}

Figures~\ref{fig:spitzer_aeiH} and~\ref{fig:spitzer_alb_rh_alpha}
plot albedo as a function of various orbital parameters. No trends are 
readily apparent between albedo and any of these parameters. 

\subsection{\textit{Effects of Completeness}}\label{sec:eocmp}

There are two possible sources of completeness that affect the 
interpretation of the albedo vs. diameter relationship derived from the 
Hilda asteroid sample, IR completeness and optical completeness. The 
asteroids in the Hilda sample (Table~\ref{table:flux_table}) with the three
highest albedos, 2003 SB45, 2005 EC205, 2003 WD111, are also the faintest 
objects in the \textit{Spitzer} 24~\micron{} sample. From the MIPS Instrument 
Handbook\footnote{http://ssc.spitzer.caltech.edu/mips/mipsinstrumenthandbook/},
the $1~\sigma$ sensitivity limit for a 42~sec. astronomical observing 
request  (AOR) is 214~$\mu$Jy in a region with high background levels such 
as those found along the line of sight through the zodical light. The 
sensitivity limit of the MIPS instrument is proportional to the square root 
of the exposure time in the background limited case. Following this latter 
convention, the $5~\sigma$ detection limit for a 3~sec. BCD frame is 
4004~$\mu$Jy. Nine asteroids are not detected to a $5~\sigma$ level in 
individual BCDs. Even if these objects are excluded from the Hilda 
population sample, there is still a trend towards higher albedo with 
smaller diameter with a correlation significance of 
4.8~$\sigma$ as determined from the Spearman rank-order test. 

The optical completeness for the Hilda asteroid group and the photometric 
uncertainty associated with the optical data may also bias the derived 
results. To assess the completeness of asteroid surveys, we assume that 
they are complete to a V magnitude of 21.5, commensurate with the 
completeness limits of the Sloan Digital Sky Survey \citep{ivezic01}, the 
Sub-Kilometer Asteroid Diameter Survey \citep{gladman09}, and Spacewatch 
\citep{larsen07}. Assuming that an asteroid will be detected at opposition 
by one of a number of surveys, the relation 
$m_{V}=H + 5 log [r_{h}(r_{h}-1)],$ can be used to estimate completeness, 
where we substituting the mean aphelion distance of 4.79 AU derived from 
the orbital parameters of 3509 known Hilda group asteriods for the 
heliocentric distance, $r_{h}$. We find that optical surveys must be 
complete to at least H$= 15.21$~magnitude. Only two objects in 
Table~\ref{table:flux_table}, 2004 TP256 and 2001 UY149, are at this 
completeness limit or a fainter limit; therefore, our Hilda sample is not 
dominated by the effects of optical completeness. The albedo and albedo 
uncertainties in Table~\ref{table:solutions_table} are derived solely from 
the uncertainties in the mid-IR photometry and do not include uncertainties 
in the H magnitudes. A $\pm 0.1$ magnitude uncertainty in H results in a 
$\simeq 10$\% variation in the value calculated albedo. Thus, the marked 
change in slope of albedo-diameter relation evident in 
Figs.~\ref{fig:all_hildas_alb_diam_scatter} and 
\ref{fig:spitzer_alb_diam_scatter} is likely significant and not an 
artifact related to sample size or optical completeness effects as the 
variation is a $\gtsimeq 10$\% deviation from the derived mean geometric 
albedo. 

\subsection{\textit{Selection of NEATM Beaming Parameter}}\label{sec:neatm_select}

The selection of a fixed value for $\eta$ in NEATM may introduce a 
systematic uncertainty in the derived values for the albedo and diameter. 
In particular, a linear relation between $\eta$ and the phase angle has 
been noted by \citet{Delbo07} and \citet{Wolters08} within the Near Earth 
Asteroid (NEA) population.  As the IRAS and \textit{Spitzer} Hilda asteroid 
samples used in our analysis contain a range of phase angles between 
$11^{\circ}$ and $18^{\circ}$, no adjustment to the 
mean beaming parameter, $\eta = 0.91$ (see \S~\ref{sec:therm_mods}) was 
used nor needed. 

The selection of a mean beaming parameter may introduce some error to the fitted albedo and diameter values for these objects. The mean beaming parameter of the large Hilda population is $\eta = 0.91$; however, the standard deviation of the beaming parameter for this asteroid population is $= 0.14$. Though this variation is smaller than the standard deviation for the main belt as a whole (where the standard deviation of $ \eta \sim 0.27$), this variation can introduce errors in the albedo and diameter calculations presented in \S~\ref{sec:therm_mods}. To characterize the range of uncertainty we re-analyzed our photometry from Table~\ref{table:flux_table} with NEATM utilizing fixed $\eta$ values of 0.77 and 1.05. For a value of $\eta=0.77$, the calculated albedos of asteroids are on average 4\% higher and the calculated diameters are 1\% smaller than the values calculated when $\eta=0.91$. If a value of $\eta = 1.05$ is used, the calculated albedos of asteroids are on aver 20\% lower and diameters increase by 12\%. This increase in beaming parameter subsequently reduces a mean small Hilda asteroid population albedo to $p_{V}= 0.053$ which is closer to the mean value for the large Hilda asteroids observed by IRAS. An increase in the beaming parameter would correspond to an increased thermal inertia, which to first order is an inverse function asteroid diameter resulting trend for the Hildas that is self-consistent with that derived for the main belt and Near Earth Asteroid by \citet{delbtan09}. However, an increase in thermal inertia would require that small Hilda asteroids retain less regolith than large Hildas, which has not yet been observed via mid-infrared spectroscopic surveys. Studies of similar primitive type objects such as small Trojan asteroids \citep{fernandez2009} and comet nuclei at 4-5 AU  \citep{fernandez2008} have derived beaming parameters near the value of $\eta=0.91$ indicating that the value used in our study of Hildas is appropriate for primitive outer solar system objects.

\subsection{\textit{The Size-Frequency Distribution}\label{sec:wds_sfd}}

The size-frequency distribution of the Hilda asteroids can be inspected to 
determine if the group is in collision equilibrium. The modeling of 
\citet{Dohnanyi1969} indicates that an asteroid population is in collisional 
equilibrium if the cumulative size-frequency distribution is near a 
diameter slope of $-2.5$. 

\begin{figure}
\epsscale{1.0}
\plotone{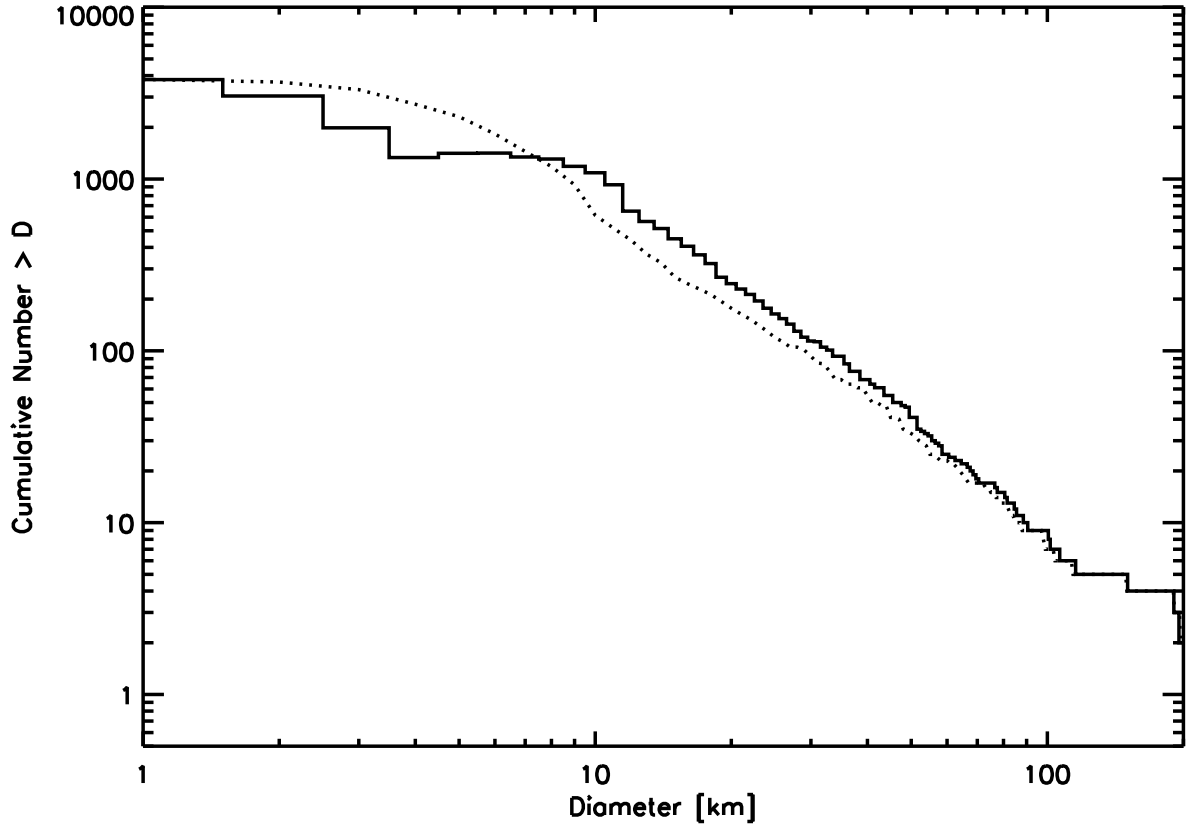}
\caption{Cumulative size frequency of Hilda asteroid group utilizing the 
albedo function as derived from our \textit{Spitzer} MIPS 24~\micron{} 
data (solid line) and an albedo of 0.04 (dotted line) corresponding 
to the average from the large Hildas detected with IRAS \citep{ryan2010}.
\label{fig:sfd}}
\end{figure}

We derived a best-fit relationship between albedo and diameter using a 
second-order polynomial of the form 

\begin{equation}
p_{V}=0.04-0.49 \left( \frac{1 km}{D} \right)+4.14 \left( \frac{1km}{D} \right)^{2} \label{eqn:alb_as_func_diam}
\end{equation}

\noindent to determine if the Hildas are in collisional equilibrium, as the 
relationship appears to be an inverse function of diameter 
(Figs.~\ref{fig:all_hildas_alb_diam_scatter}, 
\ref{fig:spitzer_alb_diam_scatter}). For all asteroid diameters $\ltsimeq 
4$~km, which are beyond the completeness limit of our sample, $p_{V}$ is 
held fixed at a value of 0.2. The albedo function described by 
Eqn.~\ref{eqn:alb_as_func_diam} was used in conjunction with H magnitudes 
of all known Hildas to derive a size-frequency distribution, 
Fig.~\ref{fig:sfd}. To illustrate the effect of using an alternative albedo 
function, we also plot a cumulative size-frequency distribution in 
Fig.~\ref{fig:sfd} assuming all Hilda group asteroids regardless of 
diameter have a mean geometric albedo of 0.04. 
 
Functionally the two size-frequency distributions appear similar at large 
diameters, and each both appear to break at diameters, $D \ltsimeq 10$~km. 
The cumulative size-frequency distribution calculated with our albedo 
function can be described by a broken power law of the form $N_{i} = \beta 
\times D{\rm(km)}^{\alpha}$ with two components. For large Hilda group 
asteroids with $D > 12$~km, $\beta = 94317 \pm 3395$ and $\alpha = -(2.00 \pm 0.02)$
while for the smaller population, $5 < D\rm{(km)} < 12$, $\beta = 2696 \pm 189$, 
$\alpha = - (0.37 \pm 0.08)$. The value of $\alpha = -2.00$ for the larger 
Hilda asteroid group population strongly suggests that they are in 
near-collisional equilibrium \citep{Dohnanyi1969}. However, there are less 
smaller Hilda asteroids observed than expected for such a system.  If the 
cumulative number distribution for diameters $>$~12~km was continuous, one 
would predict a total number of 3775 Hildas with diameters $>$~5~km, rather 
than the 1334 Hildas observed with diameters $>$~5~km. 

An explanation for the decrement of small Hilda group asteroids in the 
size-frequency distribution illustrated in Fig.~\ref{fig:sfd} is a lack of 
optical completeness. However, the optical data is complete to H $\simeq 
15$~magnitude, corresponding to a diameter of at least 6.6~km (assuming an 
albedo of 0.04). The inversion in the size-frequency distribution begins 
near $D \simeq 10$~km, far before this small diameter break point. An 
alternative interpretation is that the shallow slope at small diameters 
traces the depletion of small objects from the Hildas. Models by 
\citet{GilHuttonBrunini2000} suggest that collisions within the Hilda 
asteroid group population from scattered objects originating in the 
Uranus-Neptune zone could produce a large number of asteroid fragments with 
relative velocities high enough to escape the resonance. Thus, the 
decrement of small Hilda group asteroids could be explained via an intense 
collisional period, such as occurred in the late Heavy Bombardment epoch in 
the early solar system followed by a long period of low collisional 
activity. The age estimate of $\gtrsim$~4~Gyr for the Hilda family 
\citep{broz2008}, a dynamical subgroup of the Hilda asteroid group (see 
\S~\ref{sec:origins}) lends credence to the later hypothesis. 

A lower limit to the total Hilda asteroid group mass 
can be derived by integrating 
over the differential size distribution, $n(r)dr$, assuming an average bulk 
density, $\rho^{ave}_{Hilda}$. The differential size distribution has two 
components, 

\begin{equation}
 n_{1}(r) dr =47186 \times \left(\frac{1\; km}{r}\right)^{3.00 \pm 0.02} dr \; \; (6 \; km \le r \le 90 \; km)
\label{eqn:diff_sfd_large} 
\end{equation}
 
\noindent and
 
\begin{equation}
 n_{2}(r) dr= 772 \times \left(\frac{1\; km}{r}\right)^{1.37 \pm 0.08} dr \; \;  ( 2.5 \; km \le r \le 6 \; km)
\label{eqn:diff_sfd_small} 
\end{equation}

\noindent where $r$ is the asteroid radius. The lower limit of Hilda 
asteroid group mass is therefore, 

\begin{equation}
M_{T}= \int_0^{r_{c}} \frac{4}{3} \pi \rho^{ave}_{Hilda} r^{3}n_{2}(r) dr + \int_{r_{c}}^{90} \frac{4}{3} \pi \rho^{ave}_{Hilda} r^{3}n_{1}(r) dr 
\end{equation}

\noindent where $n_{1}$ and $n_{2}$ are from Eqns~\ref{eqn:diff_sfd_large}
and \ref{eqn:diff_sfd_small}, $\rho^{ave}_{Hilda}= 2300$~kg~m$^{-3}$ is the
bulk density, and $r_{c}$ is the radius, 12.5~km,  at which the two 
differential size distributions are equal. We find that 
$M_{T} \approx$ 4 $\times \; 10^{19}$ kg $\approx 6 \; 
\times 10^{-6}$ M$_{Earth}$, equivalent to a 165~km radius sphere having 
the same density which would be approximately twice as large as 153~Hilda.

\subsection{Origins}\label{sec:origins}

The observed albedo variations that seem diameter dependent likely is 
indicative of the Hilda asteroid group origins and evolutionary processes. 
Variation in the relative degree of metamorphism within the parent body 
population arising from differences in the extent of internal melting, 
caused by the radioactive decay nucleotides such as $^{26}$Al or $^{60}$Fe, 
might yield a population dichotomy in the currently observed Hilda asteroid 
group \citep{Mcoy06}. Asteroids with high albedos arise perhaps from the 
most thermally altered bodies in this primordial population. However, the 
Hilda asteroid group sampled in our MIPS survey likely is in collisional 
equilibrium for asteroids with diameters $>$ 10 km and thus we do not expect to see such a clear trend of albedo with diameter as evident in our results. We also do not expect a thermally modified parent body in the region of the Hilda group as the models of \citet{grimmmcsween93} indicate that a sufficent quantity of  $^{26}$Al necessary to cause internal melt would not be accreted in planetesimal bodies of all sizes formed at heliocentric distances $\gtrsim$ 3.4 AU.

The discovery of distinct dynamical families within the Hilda group by 
\citet{broz2008} necessitates examination of whether or not variations in 
albedo observed in our sample (Table~\ref{table:solutions_table}) are 
indicative of dynamical family membership instead. Utilizing the Hierarchical Clustering Method \citep{zappala1990} which utilizes proper elements of asteroids and searches phase space for clusters of objects with similar velocities, \citet{broz2008} identify two distinct dynamical families within the Hilda obital group. Of these two dynamical 
families, Hilda and Schubart, identified by \citet{broz2008}, only the 
Schubart dynamical family contributes significantly to our Hilda asteroid 
group sample. The 16 members of the Schubart family present in our sample 
follow the same albedo trend as seen within the rest of the Hilda asteroid 
group as illustrated in Fig.~\ref{fig:spitzer_alb_diam_scatter}. Thus 
dynamical family membership cannot be delineated solely by albedo.

Alternatively, the albedo-diameter relation may instead be indicative of 
the influence of collisional processes within Hilda asteroid group.  With increased age and thus increased ion 
irradiation exposure, surfaces with organic compositions redden and the 
geometric albedo is reduced \citep[][]{Andronico87,Moroz04}. This is the reported cause for the albedo-diameter relation for small Trojan asteroids \citep{fernandez2009}. Evidence of the space weathering reddening effect is observed for Hildas in the optical \citep{dahlgren95}, 
where a trend towards D-type asteroids is apparent with decreasing diameter 
for asteroids with D $>$ 20~km. It is however problematic to infer a connection between albedo and the potential for space weathering for small diameter objects as the small Hilda asteroids seen in \citet{sloanhildas08} display a significant range of spectral slopes and thus taxonomic types in the range 12 $<$ H $<$ 16. This spectral range is characteristic of the  C-, D-, and X- taxonomic types as derived by \citet{busbinzel2002} and the range of small Hilda group asteroid albedos corresponds with the range in albedos for these taxonomic types as derived in \citet{ryan2010}. If the albedo-diameter relation is  indicative of space weathering and subsequent reddening and reduced albedo with continued exposure to solar flux, one would also expect that a clear trend of increasing spectral slope as an inverse function of absolute magnitude, which is not apparent from the work of \citet{sloanhildas08}.

Interestingly, the albedo distribution of small Hilda asteroids is quite 
similar to the albedo distribution of cold classical KBOs 
\citep{brucker09}. Cold classical KBOs are thought to be the original 
proto-Kuiper Belt population \citep{levison08}, suggesting that the Hilda 
asteroid group may be contaminated by bodies which originated in orbits 
past Uranus and Neptune. The dynamical models of \citet{levison09}, predict 
that $\sim$8$\%$ of the Hilda population can be populated by bodies which 
originated in the Kuiper Belt and were subsequently transported into the 
inner solar system via giant planet migration.  Our Hilda asteroid group 
sample contains 6 asteroids with $p_{V} >$ 0.1, an albedo range 
characteristic of cold classical KBOs and higher than the albedo range for 
standard D- and C-type asteroids. Furthermore, number counts derived from 
the 24~\micron{} MIPS mid-IR fluxes above the 5~$\sigma$ detection threshold 
imply that up to 11$\%$ of the small Hilda asteroid group population is 
contaminated with objects that may have originated within the Kuiper Belt. 

\section{CONCLUSIONS} \label{sec:concl}

We have measured the 24~\micron{} thermal emission from 62 small Hilda 
group asteroids obtained by \textit{Spitzer} MIPS instrument and have 
combined this archival data set with H magnitudes to calculate effective 
diameters and albedos. Our object sample spans a range of diameters from 3 
to 12~km which is significantly smaller than the D~$>$~30~km Hilda group 
asteroids for which albedos and diameters are available from IRAS 
observations \citet{ryan2010}. Based on our analysis of the MIPS 
photometry, we conclude: 

The measured mean albedo of our small Hilda asteroid sample is $p_{V}$ = 
0.07 $\pm$ 0.05. This albedo is higher than the mean albedo of large Hildas 
which is found to be $p_{V}$ = 0.04 $\pm$ 0.01 by \citet{ryan2010} and the 
small Hilda asteroids exhibit greater albedo diversity than the larger 
members of the same dynamical population. 

The geometric albedo increases with decreasing diameter for Hildas with 
diameters in the 4 to 12~km range. The correlation is significant to a 
5.97~$\sigma$ and addition of large Hilda asteroids from IRAS observations 
\citep{ryan2010} increases the significance of this correlation to 
6.3~$\sigma$.  Though this trend could be considered to be a result of collisional processes and a tracer of space weathering, the spectral diversity of the Hilda asteroid group complicates this interpretation and colors and/or taxonomic determinations for objects in this sample are required before any firm statements can be made regarding a collisional processing and space weathering link.

The power-law slope of the Hilda asteroid group size-frequency distribution 
breaks at $\simeq$ 12.5~km when the albedo-diameter relation for Hilda asteroids is applied. This break is found to not be an observational 
bias in optical surveys, but rather a real signature wherein only 
$\sim$~30\% of asteroids with 5~km diameters predicted from the size 
frequency distribution above 15~km are observed.  Given the low collisional probabilities within the Hilda asteroid group as a whole \citep{dell'oro2001}, it is unlikely that this depletion is due strictly to Hilda-Hilda collisions resulting in small fragments with velocities sufficient to escape the resonance and is more likely the result of early depletion of the small Hilda population such as a period of intense cometary bombardment as suggested by \citet{GilHuttonBrunini2000} or bombardment from Kuiper Belt planetesimals as suggested by \citet{levison08}. As $\sim$10\% of our Hilda asteroid group sample contains asteroids with albedos commensurate with the albedos of cold classical Kuiper Belt objects, searches for Kuiper Belt contaminants within the outer solar system should include taxonomic classification and spectroscopic follow up of these asteroid targets.

\acknowledgements E.R. and C.E.W. acknowledge support from the National 
Science Foundation grant AST-0706980 to conduct this research. 

The authors thank the efforts of an anonymous referee, whose suggestions
improved our manuscript.
{\it Facilities:} \facility{Spitzer}
            
\clearpage

%% last modified cew 14.jan.2011
%%

%% REFERENCES

\clearpage

%%%%%%%%%%%%%%%%%%%%%%%%%%%%%%%%%%%%%%%%%%%%%%%%%%%%%%%%%%%%%%%%%%%
%%%%%%%%%%%%%%%%%%%%%%%%%%%%%%%%%%%%%%%%%%%%%%%%%%%%%%%%%%%%%%%%%%%%%%%%
%
% Beginning of the table section

%%%%%%%%%%%%%%%%%%%%%%%%%%%%%%%%%%%%%
%% Table 1
%%%%%%%%%%%%%%%%%%%%%%%%%%%%%%%%%%%%%
\begin{deluxetable}{llcccccccccr}
\rotate
\tabletypesize{\tiny}
\setlength{\tabcolsep}{1pt}
\tablewidth{0pt}

\tablecaption{Orbital elements and 24 $\mu$m fluxes for selected Hildas
\label{table:flux_table}}
\tablehead{\colhead{Number} & \colhead{Name/Provisional} & \colhead{Dynamical}&\colhead{Request} & \colhead{UT}& \colhead{UT} &\colhead{Heliocentric} &\colhead{Geocentric} & \colhead{Phase} & \colhead{Absolute} & \colhead {24 $\mu$m} & \colhead{Flux}\\

\colhead{} & \colhead{Designation} & \colhead{Family} &\colhead{Key}& \colhead{Date} & \colhead{Time} &\colhead{Distance} & \colhead{Distance} & \colhead {Angle} & \colhead{Magnitude} & \colhead{Flux} & \colhead{Error}\\

\colhead{} & \colhead{} &\colhead{flag}& \colhead{} & \colhead{(yyyy-mm-dd)} & \colhead{(at start)} & \colhead{(AU)} & \colhead{(AU)} & \colhead{(deg)} & \colhead{} & \colhead{($\mu$Jy)} & \colhead{($\mu$Jy)}
}
\startdata
     21804&Vaclavneumann& G&    23241472&2008-02-16& 12:54:14&3.24&2.75&   16.91&  14.500&79050.0&  1588.5 \\
     55347&    2001SH142& G&   23241728&2008-06-27& 13:08:09&3.74&3.33&   15.08&  14.500&36310.0&   738.8 \\
     79097&     1981EC24& S&   23241984&2007-11-29& 15:33:04&4.29&3.69&   11.89&  14.500&25920.0&   534.0 \\
     89903&         Post&  G&  23242240&2008-02-16& 12:45:02&3.46&3.16&   16.64&  14.500&10710.0&   252.7 \\
     90502&      Burrati&  H&   23242496&2007-11-28& 22:51:12&4.69&4.44&   12.41&  14.500&10910.0&   253.0 \\
     99862&     Kenlevin&  H&   23242752&2008-04-15& 16:26:49&4.79&4.37&   11.43&  14.500& 6570.0&   180.0 \\
    104876&     2000HH98& G&   23243008&2008-03-14& 03:46:54&4.22&4.02&   13.69&  14.500& 3956.0&   157.3 \\
    120175&     2003KB11&  G&  23243264&2007-11-30& 05:24:38&4.97&4.59&   11.37&  14.500& 5184.0&   162.4 \\
    131502&    2001SW273& S&   23243520&2007-09-26& 01:45:52&3.29&3.02&   18.04&  14.500&50340.0&  1018.4 \\
    136835&     1997UL19&G&    23243776&2008-02-16& 21:08:06&3.61&3.35&   16.07&  14.500&22350.0&   468.4 \\
    136935&      1998QK6& G&   23244288&2007-11-30& 05:41:12&4.04&3.64&   13.99&  13.900&19940.0&   419.1 \\
     39382&  Opportunity& G&   23244544&2007-08-23& 22:04:39&3.44&3.12&   17.11&  14.600&22250.0&   466.8 \\
     58353&      1995EW4& G&   23244800&2008-01-04& 02:10:45&3.56&3.46&   16.48&  14.600&33170.0&   680.8 \\
     65989&     1998KZ12& G&   23245056&2007-10-27& 11:13:43&3.93&3.52&   14.49&  14.600&29410.0&   604.6 \\
    128254&    2003SL259& S&    23245312&2008-03-17& 08:11:46&4.96&4.88&   11.67&  14.600& 5080.0&   173.0 \\
    128858&     2004SQ20&  S&  23245568&2008-03-14& 11:55:38&4.60&4.53&   12.58&  14.600&10040.0&   247.1 \\
    134429&      1998RT5&  S&  23245824&2007-11-30& 05:16:50&3.81&3.41&   14.96&  14.500&29390.0&   603.8 \\
    134562&    1999RS177& S&   23246088&2007-08-23& 23:06:59&3.14&2.67&   18.08&  13.700&65200.0&  1313.4 \\
    145396&     2005NE53&   G& 23246336&2008-03-13& 18:14:29&4.35&3.76&   11.48&  14.600& 4661.0&   150.4 \\
    145718&     1993FT57& S&    23246592&2008-05-16& 12:48:45&3.45&2.88&   15.10&  14.600&28820.0&   590.7 \\
    145960&    1999XV255& G&   23246848&2007-11-29& 10:38:27&3.64&3.41&   16.18&  14.600&21480.0&   452.1 \\
     62959&     2000VV39& G&   23247104&2007-11-28& 22:42:01&3.60&2.99&   14.23&  14.700&19930.0&   417.7 \\
     83877&     2001UE96&  G&  23247360&2007-10-27& 08:22:56&4.67&4.07&   10.94&  14.700& 9674.0&   229.3 \\
    116489&     2004BN12&  G&  23247616&2007-10-27& 11:05:38&4.14&3.75&   13.80&  14.700&14440.0&   316.7 \\
    116512&     2004BN38&  G&  23247872&2007-10-27& 11:21:22&4.44&4.04&   12.74&  14.700& 5759.0&   172.2 \\
    117667&    2005EC205& G&   23248128&2007-10-28& 06:46:22&4.43&4.09&   13.04&  14.700& 1967.0&   143.5 \\
    118177&     1992EZ13&  G&  23248384&2008-02-17& 12:20:07&3.65&3.15&   14.81&  14.300&54210.0&  1093.3 \\
    120761&      1998AX1& S&   23248640&2008-03-14& 14:37:51&4.55&4.57&   12.63&  14.700& 8685.0&   228.9 \\
    125130&     2001UO56& G&   23248896&2008-06-27& 13:16:37&3.28&2.92&   17.61&  14.100&19700.0&   415.2 \\
    128295&    2003WD111&  G&  23249152&2008-04-15& 17:16:58&4.46&4.10&   12.57&  14.700& 1764.0&   128.9 \\
    129241&     2005QS13&  S&  23249408&2007-11-29& 10:29:11&4.54&4.12&   12.31&  14.700&12200.0&   276.0 \\
    129634&     1998HP43&  G&  23249664&2007-10-27& 11:29:26&3.80&3.36&   14.84&  14.000&22090.0&   461.1 \\
    131481&    2001RT111&  G&  23249920&2007-08-27& 11:27:31&3.91&3.65&   15.10&  14.600&12330.0&   281.6 \\
    145368&     2005MB43&   G& 23250176&2007-11-30& 10:14:48&3.99&3.54&   13.94&  14.700&16820.0&   360.3 \\
     63491&     2001OY60& S&   23250432&2008-04-15& 17:08:10&3.73&3.46&   15.44&  14.800&36650.0&   747.6 \\
     64823&    2001XO240&  G&  23250688&2008-08-30& 20:35:04&3.35&3.33&   17.65&  14.800&12540.0&   291.7 \\
     73769&     1994PN12&  S&  23250944&2008-03-14& 12:10:18&4.67&4.59&   12.41&  14.800& 2657.0&   149.7 \\
     79096&     1981EM20& H&    23251200&2008-07-29& 12:50:03&4.49&4.21&   12.95&  14.800&13780.0&   306.9 \\
     85142&     1981EO29&   G& 23251456&2007-11-28& 06:58:57&4.44&3.98&   12.41&  14.800&13830.0&   302.7 \\
    119904&      2002EX6&   G& 23251712&2007-11-29& 12:20:27&4.62&4.21&   12.14&  14.100&10500.0&   243.9 \\
    129002&    2004TR256& H&   23251968&2008-04-15& 17:25:32&4.50&4.04&   12.03&  14.800& 5839.0&   168.2 \\
    133559&     2003UZ10&  G&  23252224&2008-04-15& 16:44:22&5.05&4.72&   11.17&  14.800& 2364.0&   137.4 \\
    134233&     2005YD54&   G& 23252480&2008-03-14& 03:24:36&4.48&4.19&   12.73&  14.800& 4226.0&   152.6 \\
    141518&    2002EB136& H&   23252736&2007-11-29& 12:02:25&4.35&3.79&   12.00&  14.100&21750.0&   452.6 \\
    141701&     2002KM15& S&   23252992&2007-11-29& 21:15:12&4.16&4.00&   14.15&  14.800& 6767.0&   191.9 \\
    145397&     2005NC54&   G& 23253248&2008-03-13& 10:52:27&4.37&3.80&   11.66&  14.800& 6766.0&   180.9 \\
    145421&     2005PD19& S&    23253504&2007-11-29& 21:24:34&4.31&3.95&   13.27&  14.800&13400.0&   298.7 \\
     52079&     2002RU61&  G&  23253760&2008-03-14& 03:56:04&4.15&4.15&   13.90&  14.900&17270.0&   376.6 \\
     62489&    2000SS223& S&    23254016&2007-11-28& 17:01:12&4.09&3.57&   13.16&  14.900&27430.0&   564.7 \\
    127519&     2002UJ16&  G&  23254272&2008-04-15& 16:35:41&3.86&3.52&   14.76&  14.300&18490.0&   393.0 \\
    143658&     2003SB45&  G&  23254528&2008-04-15& 16:52:11&4.94&4.63&   11.48&  14.900& 1114.0&   130.9 \\
    145373&     2005MV49&   G& 23254784&2007-11-30& 10:23:43&3.76&3.43&   15.42&  14.900&14100.0&   312.2 \\
    133324&     2003SY90&   G& 23255040&2008-03-14& 04:04:36&4.77&4.71&   12.12&  15.000& 6020.0&   186.2 \\
    145767&       1997PW&   G& 23255296&2007-11-30& 05:32:48&4.68&4.25&   11.92&  15.000& 3756.0&   145.8 \\
    134016&     2004VU53&  G&  23255552&2008-04-15& 17:00:16&4.94&4.68&   11.56&  15.100& 2942.0&   143.6 \\
    134690&     1999XP61&   G& 23255808&2008-03-13& 10:31:16&3.48&2.96&   15.37&  14.300&48220.0&   974.4 \\
    143621&     2003GE55&   G&  23256064&2007-11-29& 12:11:07&4.72&4.15&   10.95&  14.500&11720.0&   265.6 \\
     64390&    2001UY149&   G&  23256320&2007-11-28& 23:00:48&4.39&4.15&   13.32&  15.200& 9007.0&   225.1 \\
    145841&     1998YH20&   G&  23256576&2008-02-17& 12:10:41&3.83&3.57&   15.09&  15.000&15470.0&   337.2 \\
    129007&    2004TP296&   G&  23256832&2008-03-14& 03:38:42&4.99&4.83&   11.57&  15.500& 3147.0&   149.9 \\
    142470&     2002TE13&   G&  23257088&2007-08-22& 12:28:12&4.54&4.49&   13.01&  14.900& 7620.0&   209.0 \\
    147836&    2005TN125& S&    23244032&2009-03-23& 10:59:29&3.59&3.43&   16.21&  14.500&30580.0&   629.0\\
\enddata
\tablecomments{ In column 3 the flags are the following: H= member of the Hilda dynamical family as determined by \citet{broz2008}, S= member of the Shubart dynamical family as determined by \citet{broz2008} and G= member of Hilda group without either Hilda or Schubart family correspondence}
\end{deluxetable}

\clearpage

\begin{deluxetable}{lcccr}
\tablecolumns{5} 
\tablewidth{0pc} 
%\rotate
\tablecaption{MIPS 24 $\mu$m HILDA ASTEROID GROUP THERMAL MODEL SOLUTIONS
\label{table:solutions_table}}
\tablehead{
\colhead{Name} & \colhead{STM} & \colhead{STM} & \colhead{NEATM} & \colhead{NEATM}\\
\colhead{} & \colhead{Albedo} & \colhead{Diameter} & \colhead{Albedo} & \colhead{Diameter   }\\
\colhead{} & \colhead{} & \colhead{(km)} & \colhead{} & \colhead{(km)}
}
\startdata
Vaclavneumann &  0.025$\pm$0.000 &  10.59$\pm$0.06 &  0.024$\pm$0.000 &  10.73$\pm$0.06 \\
    2001SH142 &  0.061$\pm$0.001 &   9.61$\pm$0.05 &  0.058$\pm$0.001 &   9.81$\pm$0.06 \\
     1981EC25 &  0.024$\pm$0.000 &  11.35$\pm$0.07 &  0.023$\pm$0.001 &  11.64$\pm$0.28 \\
         Post &  0.191$\pm$0.003 &   4.76$\pm$0.03 &  0.185$\pm$0.003 &   4.84$\pm$0.03 \\
      Burrati &  0.090$\pm$0.001 &   7.94$\pm$0.05 &  0.084$\pm$0.001 &   8.22$\pm$0.05 \\
     Kenlevin &  0.079$\pm$0.001 &   6.41$\pm$0.03 &  0.073$\pm$0.001 &   6.66$\pm$0.05 \\
     2000HH98 &  0.181$\pm$0.003 &   4.24$\pm$0.04 &  0.171$\pm$0.003 &   4.36$\pm$0.03 \\
     2003KB11 &  0.081$\pm$0.001 &   6.10$\pm$0.04 &  0.075$\pm$0.001 &   6.34$\pm$0.04 \\
    2001SW273 &  0.039$\pm$0.000 &   9.37$\pm$0.06 &  0.038$\pm$0.000 &   9.48$\pm$0.06 \\
     1997UL19 &  0.052$\pm$0.001 &   7.38$\pm$0.05 &  0.050$\pm$0.001 &   7.51$\pm$0.04 \\
      1998QK6 &  0.042$\pm$0.001 &   8.13$\pm$0.06 &  0.040$\pm$0.001 &   8.35$\pm$0.05 \\
  Opportunity &  0.058$\pm$0.001 &   6.66$\pm$0.04 &  0.056$\pm$0.001 &   6.75$\pm$0.04 \\
      1995EW4 &  0.048$\pm$0.001 &   9.20$\pm$0.07 &  0.046$\pm$0.001 &   9.35$\pm$0.06 \\
     1998KZ12 &  0.092$\pm$0.001 &   9.46$\pm$0.06 &  0.088$\pm$0.001 &   9.69$\pm$0.06 \\
    2003SL259 &  0.067$\pm$0.001 &   6.45$\pm$0.05 &  0.062$\pm$0.001 &   6.70$\pm$0.06 \\
     2004SQ20 &  0.040$\pm$0.000 &   7.91$\pm$0.04 &  0.038$\pm$0.000 &   8.18$\pm$0.05 \\
      1998RT5 &  0.032$\pm$0.000 &   8.91$\pm$0.05 &  0.031$\pm$0.000 &   9.10$\pm$0.07 \\
    1999RS177 &  0.031$\pm$0.000 &   9.14$\pm$0.05 &  0.030$\pm$0.000 &   9.23$\pm$0.05 \\
     2005NE53 &  0.125$\pm$0.002 &   4.36$\pm$0.04 &  0.117$\pm$0.001 &   4.51$\pm$0.03 \\
     1993FT57 &  0.054$\pm$0.001 &   6.93$\pm$0.04 &  0.052$\pm$0.001 &   7.06$\pm$0.04 \\
    1999XV255 &  0.052$\pm$0.001 &   7.41$\pm$0.05 &  0.050$\pm$0.001 &   7.55$\pm$0.05 \\
     2000VV39 &  0.114$\pm$0.001 &   6.19$\pm$0.04 &  0.109$\pm$0.001 &   6.33$\pm$0.04 \\
     2001UE96 &  0.091$\pm$0.001 &   7.15$\pm$0.04 &  0.084$\pm$0.001 &   7.43$\pm$0.05 \\
     2004BN12 &  0.067$\pm$0.001 &   7.31$\pm$0.05 &  0.064$\pm$0.001 &   7.52$\pm$0.04 \\
     2004BN38 &  0.111$\pm$0.001 &   5.21$\pm$0.03 &  0.104$\pm$0.001 &   5.38$\pm$0.03 \\
    2005EC205 &  0.257$\pm$0.007 &   3.16$\pm$0.04 &  0.242$\pm$0.006 &   3.25$\pm$0.04 \\
     1992EZ13 &  0.029$\pm$0.000 &  10.86$\pm$0.08 &  0.027$\pm$0.001 &  11.13$\pm$0.24 \\
      1998AX1 &  0.046$\pm$0.001 &   7.33$\pm$0.06 &  0.043$\pm$0.001 &   7.57$\pm$0.05 \\
     2001UO56 &  0.072$\pm$0.001 &   5.68$\pm$0.03 &  0.070$\pm$0.001 &   5.76$\pm$0.03 \\
    2003WD111 &  0.240$\pm$0.005 &   3.25$\pm$0.03 &  0.225$\pm$0.004 &   3.35$\pm$0.03 \\
     2005QS13 &  0.042$\pm$0.001 &   7.86$\pm$0.05 &  0.040$\pm$0.000 &   8.13$\pm$0.05 \\
     1998HP43 &  0.041$\pm$0.001 &   7.58$\pm$0.05 &  0.039$\pm$0.000 &   7.75$\pm$0.05 \\
    2001RT111 &  0.059$\pm$0.001 &   6.29$\pm$0.04 &  0.056$\pm$0.001 &   6.43$\pm$0.03 \\
     2005MB43 &  0.057$\pm$0.001 &   7.20$\pm$0.04 &  0.054$\pm$0.001 &   7.39$\pm$0.04 \\
     2001OY60 &  0.039$\pm$0.000 &  10.00$\pm$0.06 &  0.038$\pm$0.000 &  10.19$\pm$0.06 \\
    2001XO240 &  0.137$\pm$0.002 &   5.32$\pm$0.03 &  0.133$\pm$0.002 &   5.39$\pm$0.03 \\
     1994PN12 &  0.153$\pm$0.003 &   4.26$\pm$0.04 &  0.143$\pm$0.003 &   4.41$\pm$0.05 \\
     1981EM20 &  0.054$\pm$0.001 &   8.51$\pm$0.06 &  0.051$\pm$0.001 &   8.79$\pm$0.06 \\
     1981EO29 &  0.057$\pm$0.001 &   8.07$\pm$0.04 &  0.054$\pm$0.001 &   8.33$\pm$0.05 \\
      2002EX6 &  0.037$\pm$0.000 &   7.57$\pm$0.04 &  0.034$\pm$0.000 &   7.85$\pm$0.04 \\
    2004TR256 &  0.083$\pm$0.001 &   5.33$\pm$0.04 &  0.078$\pm$0.001 &   5.52$\pm$0.04 \\
     2003UZ10 &  0.110$\pm$0.003 &   4.43$\pm$0.05 &  0.102$\pm$0.002 &   4.62$\pm$0.05 \\
     2005YD54 &  0.106$\pm$0.002 &   4.71$\pm$0.04 &  0.100$\pm$0.001 &   4.87$\pm$0.04 \\
    2002EB136 &  0.024$\pm$0.000 &   9.37$\pm$0.07 &  0.023$\pm$0.000 &   9.59$\pm$0.03 \\
     2002KM15 &  0.081$\pm$0.001 &   5.38$\pm$0.04 &  0.077$\pm$0.001 &   5.53$\pm$0.04 \\
     2005NC54 &  0.074$\pm$0.001 &   5.25$\pm$0.03 &  0.069$\pm$0.001 &   5.43$\pm$0.03 \\
     2005PD19 &  0.045$\pm$0.000 &   7.61$\pm$0.04 &  0.043$\pm$0.001 &   7.84$\pm$0.05 \\
     2002RU61 &  0.053$\pm$0.001 &   8.86$\pm$0.05 &  0.050$\pm$0.001 &   9.11$\pm$0.05 \\
    2000SS223 &  0.058$\pm$0.001 &   9.48$\pm$0.06 &  0.055$\pm$0.001 &   9.74$\pm$0.07 \\
     2002UJ16 &  0.036$\pm$0.000 &   7.37$\pm$0.05 &  0.034$\pm$0.000 &   7.54$\pm$0.05 \\
     2003SB45 &  0.316$\pm$0.013 &   2.89$\pm$0.06 &  0.294$\pm$0.013 &   2.99$\pm$0.07 \\
     2005MV49 &  0.051$\pm$0.001 &   6.12$\pm$0.04 &  0.049$\pm$0.001 &   6.24$\pm$0.04 \\
     2003SY90 &  0.059$\pm$0.001 &   6.59$\pm$0.04 &  0.055$\pm$0.001 &   6.83$\pm$0.05 \\
       1997PW &  0.075$\pm$0.001 &   4.54$\pm$0.04 &  0.070$\pm$0.001 &   4.69$\pm$0.03 \\
     2004VU53 &  0.079$\pm$0.002 &   4.66$\pm$0.04 &  0.073$\pm$0.001 &   4.84$\pm$0.04 \\
     1999XP61 &  0.039$\pm$0.000 &   9.30$\pm$0.04 &  0.038$\pm$0.000 &   9.47$\pm$0.05 \\
     2003GE55 &  0.025$\pm$0.000 &   7.96$\pm$0.05 &  0.023$\pm$0.001 &   8.34$\pm$0.24 \\
    2001UY149 &  0.071$\pm$0.001 &   6.62$\pm$0.04 &  0.067$\pm$0.001 &   6.82$\pm$0.04 \\
     1998YH20 &  0.032$\pm$0.000 &   6.75$\pm$0.03 &  0.031$\pm$0.000 &   6.90$\pm$0.04 \\
    2004TP296 &  0.062$\pm$0.001 &   5.06$\pm$0.05 &  0.057$\pm$0.001 &   5.26$\pm$0.05 \\
     2002TE13 &  0.024$\pm$0.000 &   6.75$\pm$0.05 &  0.023$\pm$0.001 &   6.95$\pm$0.11 \\
    2005TN125 &  0.042$\pm$0.001 &   8.93$\pm$0.06 &  0.041$\pm$0.000 &   9.10$\pm$0.05 \\
\enddata

\end{deluxetable}

\clearpage

%%%%%%%%%%%%%%%%%%%%%%%%%%%%%%%%%%%%%%%%%%%%%%%%%%%%%%%%%%%%%%%%%%%%%%%%
%
% Beginning of the figure section
%
% cew tweeks captions slightly 20110114

%%%%%%%%%%%%%%%%%%%%%%%%%%%%%%%%%%%%%
%% FIG 1
%%%%%%%%%%%%%%%%%%%%%%%%%%%%%%%%%%%%%

\clearpage

%%%%%%%%%%%%%%%%%%%%%%%%%%%%%%%%%%%%%
%% FIG 2
%%%%%%%%%%%%%%%%%%%%%%%%%%%%%%%%%%%%%

\clearpage

%%%%%%%%%%%%%%%%%%%%%%%%%%%%%%%%%%%%%
%% FIG 3
%%%%%%%%%%%%%%%%%%%%%%%%%%%%%%%%%%%%%

\clearpage

%%%%%%%%%%%%%%%%%%%%%%%%%%%%%%%%%%%%%
%% FIG 4
%%%%%%%%%%%%%%%%%%%%%%%%%%%%%%%%%%%%%

%%%%%%%%%%%%%%%%%%%%%%%%%%%%%%%%%%%%%
%% FIG 5
%%%%%%%%%%%%%%%%%%%%%%%%%%%%%%%%%%%%%

%%%%%%%%%%%%%%%%%%%%%%%%%%%%%%%%%%%%%
%% FIG 6
%%%%%%%%%%%%%%%%%%%%%%%%%%%%%%%%%%%%%

\end{document}